\documentclass[reprint,amsmath,amssymb,aps]{revtex4-1}

\usepackage{graphicx} 
\usepackage{dcolumn}  
\usepackage{bm} 
\usepackage{color}
\usepackage[normalem]{ulem}
\usepackage{here}

\newcommand{\beginsupplement}{%
        \setcounter{table}{0}
        \renewcommand{\thetable}{S\arabic{table}}%
        \setcounter{figure}{0}
        \renewcommand{\thefigure}{S\arabic{figure}}%
     }

\begin{document}

\title{Hexagonal Nanopits with the Zigzag Edge State on Graphite Surfaces Synthesized by Hydrogen-Plasma Etching}

\author{T. Matsui$^1$}
 \email{matsui@phys.s.u-tokyo.ac.jp}
\author{H. Sato$^1$, K. Kita$^1$, A. E. B. Amend$^1$}
\author{Hiroshi Fukuyama$^{1,2}$}
 \email{hiroshi@phys.s.u-tokyo.ac.jp}
 \affiliation{$^1$Department of Physics, The University of Tokyo, 7-3-1 Hongo, Bunkyo-ku, Tokyo 113-0033, Japan}
 \affiliation{$^2$Cryogenic Research Center, The University of Tokyo, 2-11-16 Yayoi, Bunkyo-ku, Tokyo 113-0032, Japan}

\date{\today}


\begin{abstract}

We studied,  by scanning tunneling microscopy, the morphology of nanopits of monolayer depth created at graphite surfaces by hydrogen plasma etching under various
conditions such as H$_2$ pressure, temperature, etching time, and RF power of the plasma generation.
In addition to the known pressure-induced transition of the nanopit morphology, we found a sharp temperature-induced transition from many small rather round nanopits of $\sim$150~nm size to few large hexagonal ones of 300--600~nm within a narrow temperature range.
The remote and direct plasma modes switching mechanism, which was proposed to explain the pressure-induced transition, is not directly applicable to this newly found transition.
Scanning tunneling spectroscopy (STS) measurements of edges of the hexagonal nanopits fabricated at graphite surfaces by this method show clear signatures of the peculiar electronic state localized at the zigzag edge (edge state), {\it i.e.}, a prominent peak near the Fermi energy accompanied by suppressions on either side in the local density of states.
These observations indicate that the hexagonal nanopits consist of a high density of zigzag edges. 
The STS data also revealed a domain structure of the edge state in which the electronic state varies over a length scale of $\sim 3$~nm along the edge.
The present study will pave the way for microscopic understanding of the anisotropic etching mechanism and of spin polarization in zigzag nanoribbons which are promising key elements for future graphene nanoelectronics.

\end{abstract}

\maketitle

\section{Introduction}

Reflecting the honeycomb lattice structure, graphene has two types of edges. 
One is the zigzag edge and the other the armchair one.
Among various novel physical properties of graphene~\cite{Neto2009}, the peculiar electronic state localized at the zigzag edge, the so called ``zigzag edge state" or simply ``edge state'', is one of the most attractive ones.
Such a localized state emerges due to symmetry breaking of the two equivalent sublattices in the honeycomb structure at the zigzag edge.
Inspired by the prediction by Fujita and co-workers~\cite{Fujita1996,Nakada1996}, the edge state was experimentally found by scanning tunneling microscopy and spectroscopy (STM/S) measurements at naturally existing monatomic step-edges at highly oriented pyrolytic graphite (HOPG) surfaces~\cite{NiimiMatsuiKambaraEtAl2005, NiimiMatsuiKambaraEtAl2006} and at similar step-edges in nanographites synthesized on HOPG \cite{KobayashiFukuiEnokiEtAl2005}.
The decay length of the edge state was measured as 1.2 $\pm$ 0.2~nm~\cite{NiimiMatsuiKambaraEtAl2006}.

In principle, the zigzag edge state can be spin polarized even under infinitesimally small electron-electron interaction due to its flat band nature located near the Fermi energy ($E_\mathrm{F}$)~\cite{Fujita1996}.
In graphene nanoribbons with two parallel zigzag edges (zGNRs), the edge spins are predicted to order ferromagnetically (FM) along the edge and antiferromagnetically (AF) between the edges~\cite{Fujita1996,Lee2005,Son2006,Pisani2007}.
By applying a magnetic field~\cite{Munoz-Rojas2009} or an in-plane electric field~\cite{Son2006a}, the AF configuration will flip to the FM one.
This can be applied to carbon-based switching nanodevices because the system is expected to be semiconducting with a band gap in the AF configuration while it is metallic in the FM one~\cite{Munoz-Rojas2009,Son2006a}.

In order to experimentally test the theoretical predictions, it is crucial to realize nearly pure zigzag edges, because the spin polarized states become less stable against the non-magnetic state under edge disorders~\cite{Fernandez-Rossier2007,Bhowmick2008,Jiang2008,Yazyev2011}.
After the first observations of the edge state~\cite{NiimiMatsuiKambaraEtAl2005, NiimiMatsuiKambaraEtAl2006,KobayashiFukuiEnokiEtAl2005}, numerous trials have been made to shape graphene edges on an atomic scale in order to fabricate zGNRs with the spin-polarized edge state. 
They include reactive ion etching~\cite{Han2007, Stampfer2008, Oostinga2010},
anisotropic etching using metallic nanoparticles as catalysts~\cite{Konishi2006, Datta2008, Ci2008, Campos2009, Sugiyama2014, Li2014},
carbothermal etching~\cite{Nemes-Incze2010, Krauss2010, Oberhuber2013},
nanolithography with STM~\cite{Magda2014} and atomic force microscope (AFM) tips \cite{Puddy2011},
unzipping of carbon nanotubes~\cite{Jiao2009, Jiao2010, Xie2010, Tao2011, ZhangYazyevFengEtAl2013},
and bottom-up fabrications~\cite{Li2008, Pan2012, Wang2016, Ruffieux2016}.
However, non of them has succeeded in synthesizing zigzag edges which satisfy all the  necessary conditions: (i) quality or purity (low concentration of the armchair edge), (ii) known edge termination, and (iii) scalability or reproducibility.

Among the previous attempts, the hydrogen (H) plasma etching has apparent advantages from the view point of the requirements (ii) and (iii).
It is reported that this method can create hexagonal nanopits on graphite surfaces~\cite{Yang2010,Diankov2013,Hug2017} with edges in parallel to the zigzag direction.
In addition, the nanopits can be created in arbitrary spatial arrangements, with which one can synthesize zGNR at high densities~\cite{Yang2010,Shi2011,Yang2011,Xie2012}.
However, so far, no atomic-scale spectroscopy has been applied on the edges prepared by this method.
Thus it is not clear if it meets the requirement (i) so as to be used in ``edge-state devices''.
In addition, there are conflicts among previous experimental results on monolayer graphene exfoliated on $\mathrm{SiO_2}$.
Yang {\it et al.}~\cite{Yang2010} reported successful creation of hexagonal nanopits of uniform size, while Diankov {\it et al.}~\cite{Diankov2013} and Hug {\it et al.}~\cite{Hug2017} reported circular ones.
For a zGNR unzipped from a carbon nanotube, no nanopit creation by H-plasma etching has been found~\cite{Xie2010}.

Recently, a new insight on H-plasma etching of graphite has been brought by Hug {\it et al.}~\cite{Hug2017}.
They found that anisotropic etching, that grows hexagonal nanopits, proceeds when a graphite sample is located outside of a glow regime of the plasma; the ``remote'' plasma mode.
On the other hand, when a sample is located inside the glow, defect creation is promoted, resulting in multi-layer terraced nanopits with irregularly shaped edges; the ``direct'' plasma mode.
They demonstrated that the direct/remote mode can be switched by tuning the gas pressure. 
However, the role of etching temperature and other parameters are not known. 

In this article, we report on comprehensive studies of the morphology of nanopits created at graphite surfaces by H-plasma etching. 
The shape, depth, and growth rate of the nanopits are examined on an atomic scale with STM for samples etched under various conditions.
We found that the morphology suddenly changes from many small nanopits to few large hexagonal nanopits when the reaction temperature is increased from 450 to 500~$^\circ\mathrm{C}$.
Such a transition has not been reported before.
It is also found that the morphology changes from many irregularly-shaped multi-layer terraced nanopits to few large hexagonal nanopits on increasing the hydrogen gas pressure from 13 to 60~Pa.
This pressure variation is consistent with the previous report~\cite{Hug2017}.
Most importantly, our STS measurements have verified that the edges of the hexagonal nanopits prepared by this method are zigzag edges of high purity with the most pronounced edge-state features ever reported.
Spatial variations of the edge state studied by STS are also discussed.

\section{Results and discussion}

\subsection {Pressure Dependence}

\begin{figure*}[t]
\centering
\includegraphics[width=1.0\linewidth]{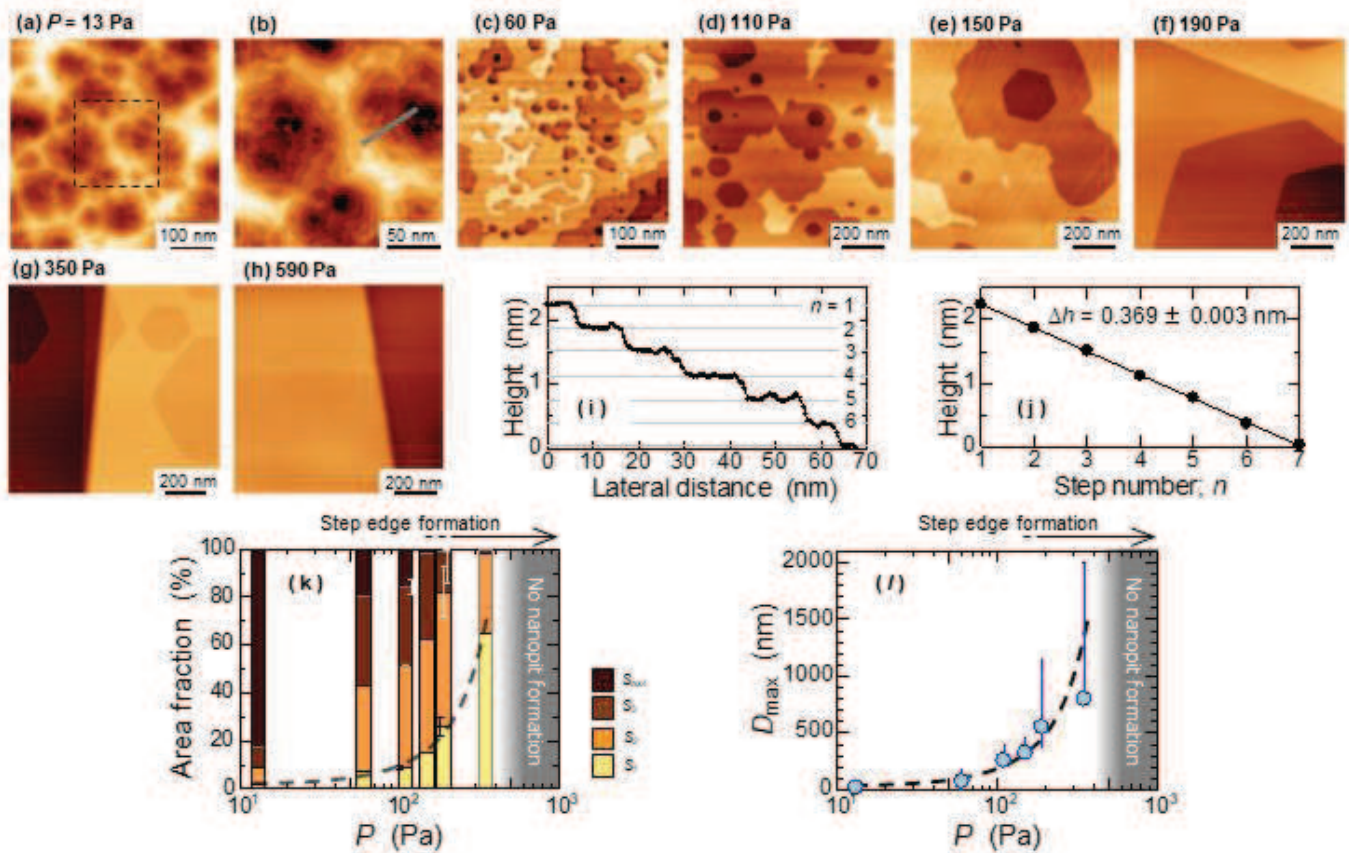}
\caption{
STM images of graphite surfaces etched by H-plasma at different $P$ of 
(a)(b) 13, 
(c) 60, 
(d) 110, 
(e) 150,
(f) 190,
(g) 350, 
and (h) 590~Pa.
The other parameters are fixed at $T = 500~^\circ\mathrm{C}$, $t = 40$~min., and $W_\mathrm{RF} = 20$~W. 
(b) Magnified image of the region indicated by the square in (a).
(i) Cross sectional profile along the solid line in (b), and (j) relation between the terrace height and the step number showing that the nanopit steps are of monolayer height. 
Note that the heights of the linear steps in (g) and (h) are much larger, {\it i.e.}, 10 and 4 layers height, respectively.
(k) $S_\mathrm{n}$ and (l) $D_\mathrm{max}$ are plotted as a functions of $P$. 
It is clear that the etching character changes rapidly between 13 and 60~Pa from irregular to anisotropic as $P$ increases, which is consistent with the observation in Ref.~\citenum{Hug2017}.
However, in our case, the size distribution of the nanopits is much wider and $D_\mathrm{max}$ is much larger.
}
\label{fig:pressure}
\end{figure*}

Let us first show the H$_2$ gas pressure ($P$) dependence of H-plasma etching, where we obtained results that agree very well with those by Hug {\it et al.}~\cite{Hug2017}.
Figures~\ref{fig:pressure}a-h are typical topographic STM images of graphite surfaces etched at several different $P$ from 13 to 590~Pa, where etching temperature ($T$), etching time ($t$), and RF power to generate plasma ($W_\mathrm{RF}$) are fixed at $500~^{\circ}\mathrm{C}$, $40~\mathrm{min.}$, and 20~W, respectively.
At the lowest pressure of 13~Pa, the surface is fully covered by irregularly-shaped nanopits with many steps down to the 10th layer from the top as shown in Figures~\ref{fig:pressure}a and b, where \ref{fig:pressure}b is a magnified image of a region in \ref{fig:pressure}a.
The cross-sectional profile along the line indicated in Figure~\ref{fig:pressure}b is shown in Figure~\ref{fig:pressure}i, and each terrace height is replotted as a function of the step number in Figure~\ref{fig:pressure}j.
From this, the averaged step height is $0.369\pm0.003$~nm, which is 10~\% larger than the interlayer distance of graphite (0.335~nm).
Possible intercalation of atomic H, one of the components of H plasma during the etching process, and its subsequent deintercalation in vacuum may expand the interlayer distance~\cite{Waqar2007,Eren2012}.
Clearly, the defect formation is very active here.

As $P$ increases to 60~Pa, the surface morphology suddenly changes, {\it i.e.}, the number of steps decreases rapidly, and the nanopit shape changes closer to hexagonal (Figure~\ref{fig:pressure}c).
At a slightly higher $P~(= 110$~Pa), all the nanopits become hexagonal and larger in size (Figure~\ref{fig:pressure}d) indicating that the defect formation is suppressed and the anisotropic lateral etching becomes dominant.
Such tendencies steadily proceed up to $P = 190$~Pa (Figures~\ref{fig:pressure}e-f) where only few very large hexagonal nanopits (500--600~nm) grow.
It is not easy to judge if the same tendencies still continue at $P \geq 150$~Pa, partly because we cannot exclude the possibility that the top most layer has already been etched away, and partly because a new structure, which consists of deep linear-step-edges, appears as can be seen by the dark regions on the left and right sides of Figures~\ref{fig:pressure}g and h, respectively.
The height of such step-edges varies randomly from monatomic to 40 layers, and their crystallographic direction is also randomly oriented with respect to the honeycomb lattice of graphite.

We analyzed these STM images quantitatively by evaluating relative terrace areas of different layers.
The results are shown in Figures~\ref{fig:pressure}k and l where $S_\mathrm{n}$ is an areal fraction of the $n$-th layer from the top and $D_\mathrm{max}$ is the diameter, for irregularly-shaped nanopits, or the diagonal length, for hexagonal ones, of the largest nanopit.
In cases where hexagonal nanopits substantially overlap, $D_\mathrm{max}$ was estimated from the length of one of six sides of the hexagon, assuming its regular shape.
Here, 0.22, 3.8, 12, 5.0, 17, 20 and 5~$\mu\mathrm{m}^2$ areas were surveyed for samples treated at $P =$ 13, 60, 110, 150, 190, 350 and 590~Pa, respectively.
In these figures, it is clear that the etching nature changes from irregular to anisotropic with increasing pressure within a narrow range between 60 and 110~Pa.
This agrees very well with the result by Hug {\it et al.}~\cite{Hug2017} who observed a similar change of etching character between 40 and 70~Pa and attributed it to the direct/remote plasma mode change.

We have checked the applicability of the direct/remote plasma scenario to the present data.
With our experimental setup, when the plasma glow edge is located $\it{inside}$ the furnace, it is not possible to measure a distance ($l_\mathrm{g}$) between the glow edge and the end of the RF coil along the quartz tube.
Thus, $l_\mathrm{g}$ in this regime was estimated from the known relation, $l_\mathrm{g} \propto 1/\sqrt{P}$~\cite{Hug2017}, whose proportionality constant was determined by measuring $P$ and $l_\mathrm{g}$ when the glow edge is located $\it{outside}$ the furnace on the downstream side.
The estimation shows that $l_\mathrm{g}$ reaches the sample position ($l_\mathrm{s} = 0.36$~m) when $P \geq 60$~Pa, which is consistent with the direct/remote scenario.

The glowing region should roughly represent the spatial extension of H-ions such as H$_1^+$, H$_2^+$, and H$_3^+$.
The life time of the H-ions shortens with increasing gas pressure because of more frequent collisions, resulting in a shorter $l_\mathrm{g}$.
On the other hand, H-radicals created by ionization and recombination processes survive beyond the glowing region.
Therefore, one can expect that the plasma species responsible for the defect creation and the isotropic etching (anisotropic etching) in the direct (remote) regime is H-ions (H-radicals)~\cite{Hug2017}.
According to the measurement of H-plasma components as a function of pressure~\cite{Nunomura2007,FeltenMcManusRiceEtAl2014}, H$_3^\mathrm{+}$ seems to be the main contributor over the other H-ions.

It should be noted, however, that there is a quantitative difference between our result and the previous one by Hug {\it et al.}~\cite{Hug2017}.
We found a large nanopit size distribution even in the remote plasma regime at $P \geq 110~\mathrm{Pa}$ suggesting that the defect-creation is active over a wide $P$ range, while they reported only hexagonal nanopits of uniform size suggesting that the nanopits grow only from preexisting defects.
This difference may come from a slightly larger oxygen contamination in our gas flow system because oxygen is known to be strongly active in creating surface defects on graphite~\cite{Kim2009}.

\subsection{Temperature Dependence}

\begin{figure*}[t]
\centering
\includegraphics[width=1.0\linewidth]{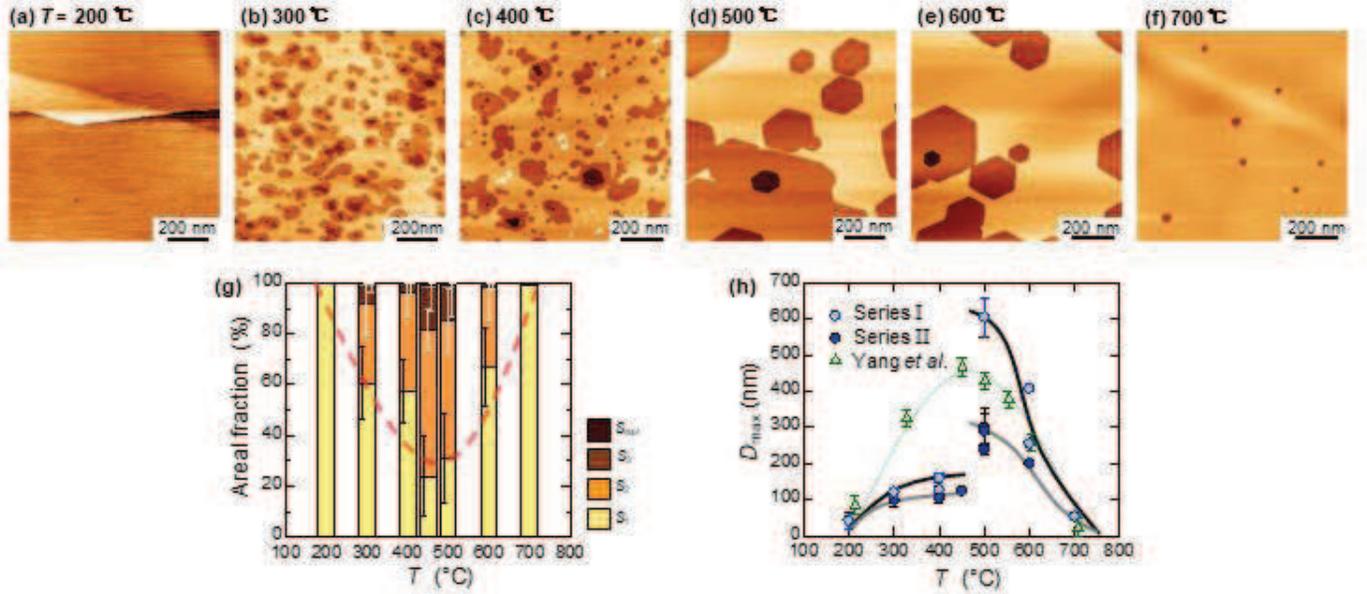}
\caption{
STM images of graphite surfaces etched by H-plasma at different $T$ of 
(a) 200, 
(b) 300, 
(c) 400, 
(d) 500, 
(e) 600, and
(f) $700~^\circ\mathrm{C}$. 
The other etching parameters are fixed at $P =$ 110~Pa, $t = 40~\mathrm{min.}$, and $W_\mathrm{RF} = 20~\mathrm{W}$. 
(g) $S_\mathrm{n}$ and (h) $D_\mathrm{max}$ are plotted as a function of $T$. The dashed line in (h) is the result by Yang {\it et al.}~\cite{Yang2010} obtained with 30~Pa, 40~min., and 20~W.
Etching characteristics drastically change between 450 and $500~^\circ\mathrm{C}$ from the more isotropic lateral etching with active defect formation to the anisotropic lateral etching with reduced defect formation. 
}
\label{fig:temperature}
\end{figure*}

Next, we examined the temperature dependence of the H-plasma etching with fixed parameters of $P = 110$~Pa, $t = 40$~min., and $W_\mathrm{RF} = 20$~W. 
Surprisingly, we found an abrupt change of the etching character within a narrow temperature range between 450 and $500~^\circ\mathrm{C}$, as will be described below.
Figures~\ref{fig:temperature}a-f show STM images of graphite surfaces etched at different $T$ from 200 to $700~^\circ\mathrm{C}$.
Here, the number of surface steps remains less than three layers at any $T$, since the etching conditions belong to the remote plasma regime in which heavy etching like Figure~\ref{fig:pressure}a never occurs.

At $T = 300~^\circ\mathrm{C}$, many small nanopits with rather round shape ($D_\mathrm{max} \approx 110$~nm) are created.
At 400 and $450~^\circ\mathrm{C}$ (not shown), the size distribution becomes wider and the shape becomes more hexagonal. $D_\mathrm{max}$ is less than 150~nm here.
On the other hand, at $500~^\circ\mathrm{C}$ (Figure~\ref{fig:temperature}d), the situation changes drastically where fewer large hexagonal nanopits are found ($D_\mathrm{max} \approx 300$ or 600~nm, depending on the measurement series as described below).
This can be seen more clearly in Figure~\ref{fig:temperature}h where $D_\mathrm{max}$ jumps by a factor of two to three across the transition temperature, $450~^\circ\mathrm{C} \leq T_\mathrm{c} \leq 500~^\circ\mathrm{C}$.
In contrast, the relative ratio $S_1$ : $S_2$ : $S_3$ ($\approx 30\%:55\%:15\%$) does not show such a jump across $T_{\mathrm{c}}$, as shown in Figure~\ref{fig:temperature}g. 
This suggests that the etching efficiency regardless of the anisotropy has a continuous dome-shaped $T$-dependence sketched by the dashed line in the figure.
At $T = 700~^\circ\mathrm{C}$ (Figure~\ref{fig:temperature}f), only tiny hexagonal nanopits ($D_\mathrm{max} \approx 50$~nm) are found at a very low density at the surface layer.

Previous researchers~\cite{Yang2010,Diankov2013} reported similar dome-shaped $T$-dependences of the etching rate but no characteristic change in the nanopit shape (see the dashed line in Figure~\ref{fig:temperature}h).
In general, etching reactions are expected to be suppressed at sufficiently low $T$ because of the lack of thermal energy.
They will also be suppressed at sufficiently high $T$ either due to the instability of CH$_4$~\cite{Wood1969}, which is expected to be the final product of the reaction~\cite{Pan1990}, or due to thermal desorption of H before etching the surface~\cite{DavydovaDespiau-PujoCungeEtAl2015}.
These are probable reasons for such a $T$ dependence.

We checked the reproducibility of the sharp transition at $T_\mathrm{c}$ by carrying out an independent series of measurements (Series II) after the first series (I).
As seen in Figure~\ref{fig:temperature}h, the two data sets agree with each other in terms of the abrupt jump in $D_\mathrm{max}$ at the same $T_\mathrm{c}$.
However, $D_\mathrm{max}$ values at $T > T_\mathrm{c}$ are two times larger in Series I than II. 
Also, in Series II, $S_3$ and $S_4$ are larger than in Series I where no $S_4$ is seen. 
The defect creation rate in Series II seems to be slightly enhanced. 
One possible explanation for this is a change of oxygen contamination in the gas flow system.

In order to test the applicability of the direct/remote mechanism to the $T$-driven mechanism found here, we have examined if $l_\mathrm{g}$ changes by varying temperature.
The glow edge location was again visually observed outside the furnace on the downstream side at $P = 13$~Pa and $W_\mathrm{RF} = 20$~W.
Eventually, we found that $l_\mathrm{g}$ changes but in the opposite direction to that expected from the model at first glance.
Specifically, $l_\mathrm{g}$ increases by 30~mm with increasing $T$ from 300 to 600~$^\circ\mathrm{C}$.
Thus some unknown mechanism may control anisotropic etching independently from the direct/remote one.
However, this cannot explain why Hug {\it et al.}~\cite{Hug2017} found the anisotropic etching even at $T (= 400~^\circ\mathrm{C}) \leq T_{\mathrm{c}}$.
Alternatively, increasing $T$ may affect the composition of the H-species in the sample region. 
For example, the fraction of H-radicals near the sample position may increase, which promotes anisotropic etching and elongates the glowing length.

In any case, to clarify the mechanism behind the the $T$-driven transitional change of the etching character we found, direct measurements of the H-plasma components in the sample region and theoretical investigations of roles of the temperature and H-radical are highly desired.

\subsection {Time Dependence}

\begin{figure}[t]
\centering
\includegraphics[width=1.0\linewidth]{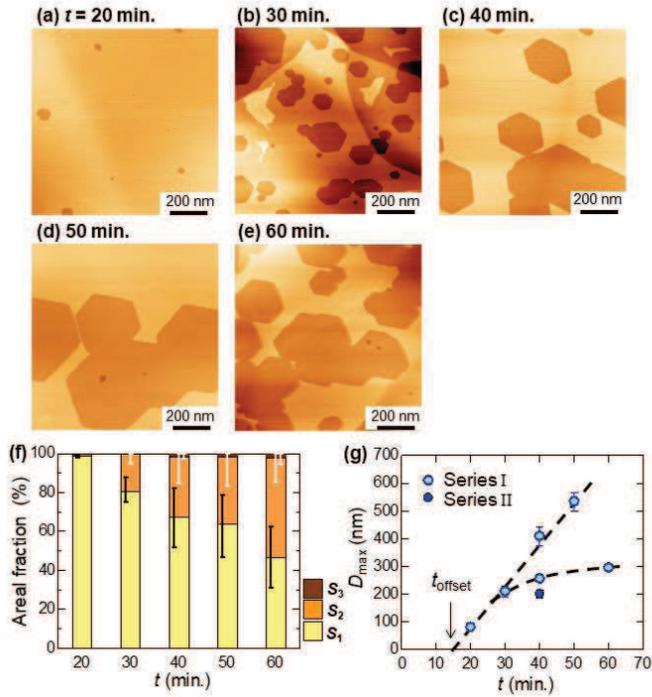}
\caption{
STM images of graphite surfaces etched by H-plasma for different $t$ of (a) 20, (b) 30, (c) 40, (d) 50, and (e) 60~min. The other parameters are fixed at $T = 600~^\circ\mathrm{C}$, $P = 110$~Pa and $W_\mathrm{RF} = 20$~W. 
(f) $S_\mathrm{n}$ and (g) $D_\mathrm{max}$ are plotted as functions of $t$.
There seems to exist a finite offset time of 10-15~min. before the anisotropic etching starts to proceed. 
}
\label{fig:time}
\end{figure}

We studied the etching time dependence as well.
Figures~\ref{fig:time}a-e show STM images of graphite surfaces etched for different $t$, at $T = 600~^\circ\mathrm{C}$, $P = 30~\mathrm{Pa}$, and $W_\mathrm{RF} = 20~\mathrm{W}$.
$S_1$ decreases linearly with increasing $t$ as expected from a constant etching rate, while $S_3$ and $S_4$ remain very small, even when $S_1 \approx 50$\% at $t = 60$~min. (see Figure~\ref{fig:time}f), because the defect formation is rather suppressed at this temperature. 

Although $D_\mathrm{max}$ data are widely scattered, there seems to exist a finite time offset of the order of 10-15~min. before the nanopits start to grow (Figure~\ref{fig:time}g). 
Nanopits in the third layer also start to appear after a longer time offset of 30-40~min.
The finite time offsets observed clearly in this inactive defect formation condition can give some insights into the kinematics of the surface defect formation.
Recent molecular dynamics simulations~\cite{DavydovaDespiau-PujoCungeEtAl2015,Despiau-PujoDavydovaCungeEtAl2016,HarpalePanesiChew2016} demonstrated that finite fluence is necessary for H ions to be adsorbed, break C-C bonds, and dissociate C atoms away on a graphene or graphite surface.

\subsection {Plasma Power Dependence}

\begin{figure}[t]
\centering
\includegraphics[width=0.97\linewidth]{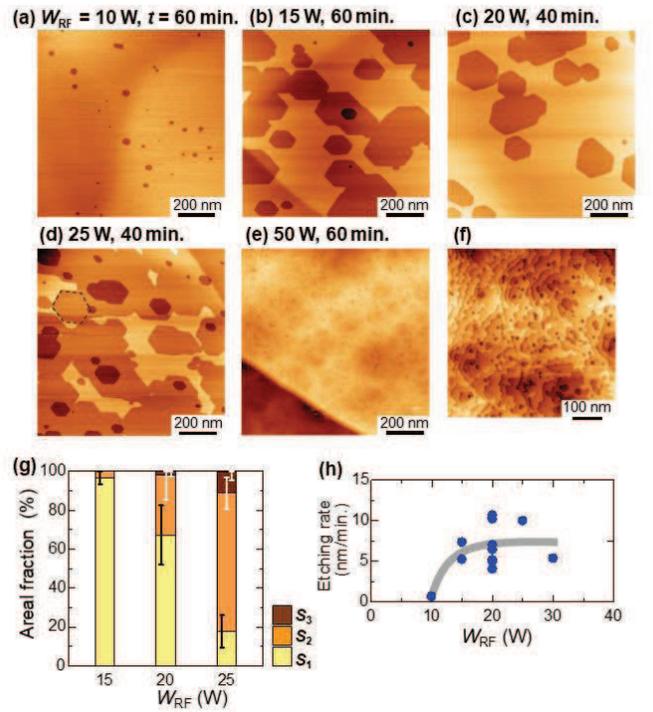}
\caption{
STM images of graphite surfaces etched by H-plasma generated by different RF powers $W_\mathrm{RF}$ of (a) 10, (b) 15, (c) 20, (d) 25, and (e)(f) 50~W. 
The etching time $t$ is (a)(b)(e)(f) 60~min. and (c)(d) 40~min.
The other etching parameters are kept at $T = 600~^\circ\mathrm{C}$ and $P = 110$~Pa. 
(g) $S_\mathrm{n}$ and (h) Etching rate are plotted as functions of $W_\mathrm{RF}$, there seems to exist a finite offset power of 10--15~W until the anisotropic etching starts.
The etching nature seems be changed between 25 and 50~W from the anisotropic lateral etching with reduced defect formation to the isotropic lateral etching with active defect formation. 
}
\label{fig:power}
\end{figure}

Figures~\ref{fig:power}a-f show STM images of graphite surfaces etched at various $W_\mathrm{RF}$ from 10 to 50~W.
Other parameters are fixed at $T = 600~^\circ\mathrm{C}$ and $P = 110$~Pa except that $t = 60$~min. for (a)(b)(e)(f) and 40~min. for (c)(d).
These parameters belong to the remote plasma mode. 
As expected, when $W_\mathrm{RF}$ increases from 10 to 25~W, the change in the hexagonal nanopit formation is more or less the same as the change caused by increasing $t$ in Figures~\ref{fig:time}a-e.
In this range, $S_1$ decreases and $S_2$ and $S_3$ increase monotonically with a finite offset of 10--15~W (see Figures~\ref{fig:power}g,h).
An interesting result was obtained at the highest RF power (= 50~W, Figures~\ref{fig:power}e,f). 
Here, the surface morphology completely changes to many multi-layer terraced nanopits with irregularly shaped edges  just like those in Figures~\ref{fig:pressure}a,b.

\begin{figure*}[t]
\centering
\includegraphics[width=0.95\linewidth]{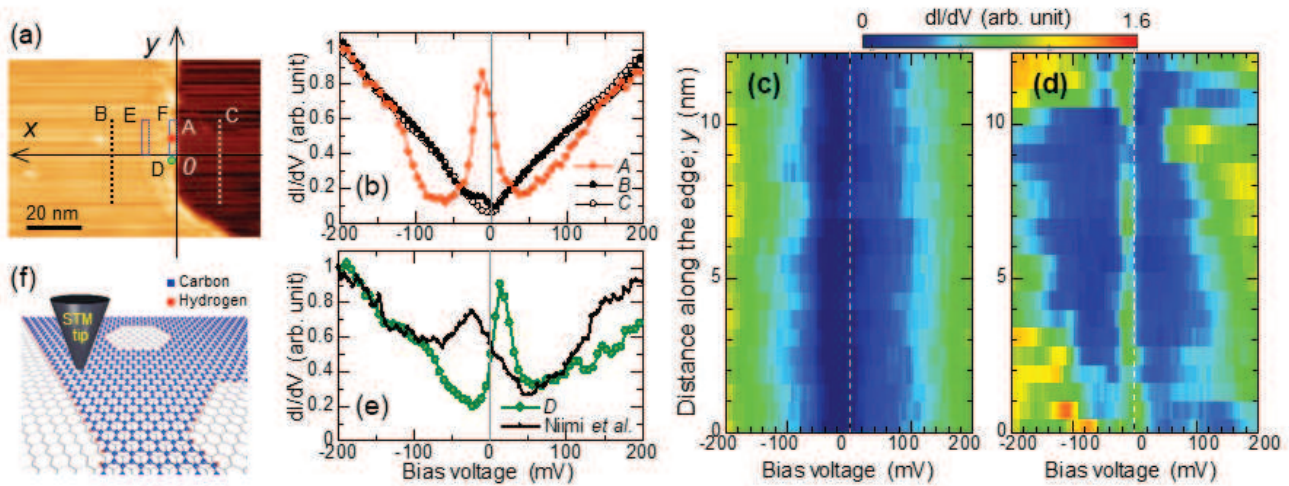}
\caption{
(a) STM image of an HOPG surface etched by H-plasma.
The right dark side shows the corner of a hexagonal nanopit with zigzag edges of monolayer height.
(b) Tunnel spectra obtained at three characteristically different positions that are near the edge (orange dots), on the upper terrace (black dots), and on the lower terrace (open circles) of the nanopit shown in (a). The specific positions are indicated in (a) by the orange dot ($x = 1$~nm, $4 \leq y \leq 7$~nm), the black dotted line ($x = 23$~nm, $-19 \leq y \leq 12.5$~nm), and the white dotted line ($x = -16$~nm, $-19 \leq y \leq 12.5$~nm), respectively.
Only near the edge, a prominent peak at $V \approx 0$ and adjacent d$I$/d$V$ suppressions are found, which correspond to the zigzag edge state.
(c)(d) Color plots of d$I$/d$V$ as a function of $y$ taken in the dotted rectangles in (a), which are at distances $x = 10$ (c) and 1~nm (d) from the edge, respectively.
The edge state exists only in (d) with a spatial variation along the edge (domain structure; see main text).
(e) Tunnel spectrum taken at a region of $x = 1$~nm and $-3 \leq y \leq 0$~nm where $V_\mathrm{peak}$ appears at a slightly positive value.
All the data were taken at $T = 4.7$~K.
The small dots are the spectrum near a zigzag step edge of monoatomic height at an exfoliated graphite surface from Ref.~\citenum{NiimiMatsuiKambaraEtAl2005}.
(f) Cartoon of hexagonal nanopits synthesized on a graphite surface by H-plasma etching and an STM tip for STS observation.
The edge carbon atoms are expected to be terminated by single hydrogen atoms.
}
\label{fig:spectroscopy}
\end{figure*}

Again, we have checked a $W_\mathrm{RF}$ dependence of $l_\mathrm{g}$ when the glow edge is located outside the furnace, and found that when $W_\mathrm{RF}$ increases from 25 to 50~W, $l_\mathrm{g}$ increases by 40--50~mm at $T = 600~^\circ\mathrm{C}$ and $P = 40$~Pa and at $T = 23~^\circ\mathrm{C}$ and $P = 14$~Pa.
Although we don't know exactly what is happening inside the furnace, it is reasonable to speculate that the observed change in surface morphology is caused when the plasma glow edge moves across the sample region with increasing $W_\mathrm{RF}$ from 25 to 50~W.

\subsection {Scanning Tunneling Spectroscopy of Zigzag Edge State at Hexagonal Nanopit Edges}

It is possible to judge if the linear edges of hexagonal nanopits are zigzag or armchair type by measuring relative angles ($\theta$) between one of atomic rows of the B-site carbon atoms, which are selectively visible with STM, and the edges~\cite{NiimiMatsuiKambaraEtAl2005, NiimiMatsuiKambaraEtAl2006,KobayashiFukuiEnokiEtAl2005,Kobayashi2006}.
The data indicate that all the edges of hexagonal nanopits synthesized by the present H-plasma etching technique are exclusively of zigzag type ($\theta = 0^\circ$, $60^\circ$, or $120^\circ$).
It is, however, technically very difficult to identify exact arrangements of the endmost carbon atoms along the edges even with the STM technique.
Instead, at present, STS observation of the edge state would be the most direct and sensitive detection of the edge quality or purity of zigzag edge structure.
This is because the edge state is known to exist in the close vicinity of only the zigzag edge and not the armchair one~\cite{Fujita1996,NiimiMatsuiKambaraEtAl2005, NiimiMatsuiKambaraEtAl2006,KobayashiFukuiEnokiEtAl2005}.
We have examined many different edges of monolayer depth with a low temperature STM/S apparatus operated at 4.7~K and 78 K in ultra high vacuum.

Figure~\ref{fig:spectroscopy}a is an STM image of one of corners of a hexagonal nanopit with a diagonal length of 160~nm created on an HOPG surface.
Etching conditions applied here are the same as those for the sample shown in Figure~\ref{fig:time}d (600~$^\circ\mathrm{C}$, 110~Pa, 50~min., 20~W).
A typical tunneling spectrum taken in the vicinity of the long vertical edge in Figure~\ref{fig:spectroscopy}a is shown by the orange circles in Figure~\ref{fig:spectroscopy}b, while those taken far from the edge are shown by the black dots (on the upper terrace) and the open circles (on the lower terrace).
The locations where the three spectra were taken are indicated by the orange dot (near edge), the black dotted line (upper terrace), and the white dotted line (lower terrace) in Figure~\ref{fig:spectroscopy}a.
Here, d$I$/d$V$, the differential tunnel conductance which is proportional to the local density of states (LDOS) of the graphite surface, is plotted as a function of the bias voltage ($V$) between the STM tip and the sample surface.
$V$ corresponds to an energy difference from the $E_\mathrm{F}$.
In Figure~\ref{fig:spectroscopy}b, only in the spectrum near the edge a prominent peak is found at $V \approx 0$, which is characteristic of the edge state.
The peak voltage ($V_\mathrm{peak}$) here is $-11$~mV.
In addition to that, the peak is accompanied by clear LDOS suppressions on both sides of the peak.
Such clear suppressions have never been reported before.
As we reported recently~\cite{Amend2017}, they should be indicative of a large sublattice imbalance~\cite{CrestiOrtmannLouvetEtAl2013, PereiraSantosNeto2008} indicating the high purity of the zigzag edge synthesized in this work.
This is the first spectroscopic evidence for the graphene zigzag edge state created by H-plasma etching.

It is noted that the tunnel spectral results are consistent with Raman spectroscopy measurements of the D band peak which we made separately for multilayer graphene samples with and without the successive anisotropic etching (LP etching) after the initial isotropic etching (HP etching). 
See Supporting Information for more details.

Next we discuss the spatial variation of the edge state studied by the STS technique particularly along the edge.
Let us define $xy$ coordinate axes with the origin $O$ as indicated in Figure~\ref{fig:spectroscopy}a.
Figures~\ref{fig:spectroscopy}c and d are color plots of d$I$/d$V$ as a function of $y$ (parallel to the edge) taken at the two different distances $x = 10$ (c) and 1~nm (d) from the edge.
The scanned areas for these STS data are indicated by the dotted rectangles in Figure~\ref{fig:spectroscopy}a, in which the d$I$/d$V$ spectra were averaged over 2~nm widths in the $x$ direction.
Since the edge state has a short decay length of only 1--2~nm~\cite{NiimiMatsuiKambaraEtAl2006,Amend2017}, one cannot detect an LDOS enhancement near $V = 0$ at all at $x = 10$~nm (see Figure~\ref{fig:spectroscopy}c).
On the other hand, in the close vicinity of the edge ($x = 1$~nm, Figure~\ref{fig:spectroscopy}d), the edge state exists almost all along the measurement range except at several positions near $y = 2, 5.5,$ and 9~nm where the LDOS peak is less clear or almost disappears. 
In other words, the edge state is divided spatially into $\sim3$~nm long domains.
While $V_\mathrm{peak}$ is unchanged within each domain, it varies randomly between domains within {$-20 \leq V_\mathrm{peak} \leq 20$~mV.
For example, the spectrum averaged over the domain located at $-3 \leq y \leq 0$~nm has a peak at $+14$~mV (the green large circles in Figure~\ref{fig:spectroscopy}e).

We observed similar domain structures for almost all edges we examined, and they may stem from their geometrical imperfections.
In any case, the edges must be highly zigzag-rich and presumably of the best quality ever achieved judging from the observed prominent LDOS peak and clear adjacent suppressions.
This can be seen by comparing it with the previously known edge state spectra~\cite{NiimiMatsuiKambaraEtAl2005,NiimiMatsuiKambaraEtAl2006}.
One of them is} shown by the small dots in Figure~\ref{fig:spectroscopy}e~\cite{NiimiMatsuiKambaraEtAl2005}.

\section {CONCLUSION}

In summary, by use of STM, we have investigated nanopits with monolayer depth created at graphite surfaces by H-plasma etching varying several key parameters such as temperature, H$_2$ pressure, etching time, and plasma generation RF power, independently.
We found a new sharp $T$-induced transition of the surface morphology from many small nanopits of rather round shape to few large hexagonal nanopits within a narrow range between 450 and 500~$^{\circ}\mathrm{C}$.
In addition, we observed a $P$-induced transition from irregularly-shaped multi-layer terraced nanopits to few large hexagonal nanopits, which is consistent with a recent report by other researchers.~\cite{Hug2017}
Our own measurement of the pressure dependence of the glowing plasma edge location supports their proposal, {\it i.e.}, the lateral etching proceeds isotropically (anisotropically) when the graphite sample is located inside (outside) of the glowing regime.
However, the $T$-induced transition we found here cannot be explained straightforwardly by this model.

Edges of the hexagonal nanopits synthesized here are always aligned to the zigzag direction and not the armchair one.
Moreover, characteristic LDOS features of the zigzag edge state, an electronic state localized at the zigzag edge, are clearly observed in STS measurements. 
This is spectroscopic evidence of the high purity zigzag edge.
Thus the anisotropic H-plasma etching is a promising route for the fabrication of zGNR, one of the key device elements for future graphene nanoelectronics. 
The STS observations also show a spatial distribution of the edge state of the order of 2--3~nm along the edge.
This demonstrates the high sensitivity of this tool to evaluate the edge quality on an atomic scale.
The present work will hopefully stimulate theoretical works to elucidate the roles of the sample temperature and H-radicals in the anisotropic H-plasma etching of graphite surface and graphene.

\subsection{Methods}

\begin{figure}[b]
\centering
\includegraphics[width=0.9\linewidth]{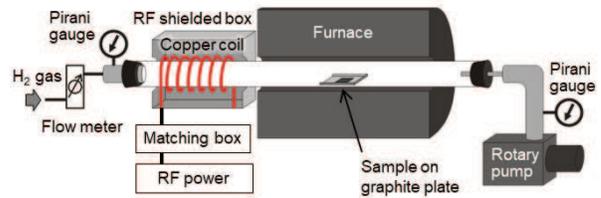}
\caption{
Schematic diagram of the experimental apparatus for H-plasma etching.
Graphite samples are located in a quartz tube and a muffle furnace under a continuous downstream of the H-plasma.
}
\label{fig:setup}
\end{figure}

Hydrogen plasma etching was performed with a home made apparatus illustrated in Figure~\ref{fig:setup}.
Freshly cleaved graphite samples ($\sim5\times2\times0.3~\mathrm{mm^3}$) are placed on a Grafoil plate (Grafoil, Graftech Inc.; 42$\times$18~mm$^2$ with 3~mm high rims at three edges) and inserted in a quartz tube (42~mm~{\it i.d.}, 1.2~m long), at the center of a muffle furnace.
The radial location is 2~mm from the tube inner edge.
We examined graphite supplied from three different companies (ZYA grade HOPG, Advanced Ceramic Corp.; PGC, Toshiba Mitsubishi-Electric Industrial Systems Corp.; Kish graphite type-B, Covalent Materials Corp.), and found no essential difference in the etching results.
H$_2$ gas (7N purity) flows at a constant rate, while keeping $P$ at the sample location constant by evacuating with a rotary vacuum pump with pumping speed 200~L/min through a needle valve.  
$P$ was estimated from pressure values measured at both ends of the quartz tube by pirani gauges and  the calculated conductance of the system. 
H-plasma is generated at 13.56~MHz with a copper coil wound around the quartz tube and an RF power source (model TX03-9001-00, ADTEC plasma Technology co., Ltd.).

After stabilizing the furnace temperature and the H$_2$ pressure, the H-plasma etching starts by turning the RF power on.
After etching for a certain time duration, the power supplies both to the plasma generator and the furnace are turned off at the same time.
It takes about 4~hours for the furnace to cool down to 200~$^{\circ}\mathrm{C}$ from 500~$^{\circ}\mathrm{C}$  while continuing to vacuum pump the quartz tube.

The morphology of etched sample surfaces is examined by two different STMs (UNISOK Co., Ltd.) operated by a commercial controller (model SPM100, RHK Technology, Inc.).
All topographic images are obtained in the constant current mode ($I = 1.0$~nA, $V = 500$~mV) in  atmospheric conditions.
STS data are taken at 4.7~K in ultra-high vacuum using a home made STM/S system which can work in multi-extreme environments~\cite{KambaraMatsuiNiimiEtAl2007}.


We thank Y. Ohba and Y. L. Liu for their technical assistances in construction of the H-plasma etching apparatus and STS data analyses.
This work was financially supported by Grant-in-Aid for Young Scientists (B) (Grant No. 25800191), Scientific Research (C) (Grant No. 15K05159), and Scientific Research (B) (Grant No. 18H01170) from JSPS.
Authors acknowledge the free-to-use software, WSxM~\cite{WSxM}.
We also appreciate the Cryogenic Research Center and the MERIT program of the University of Tokyo for their supply of liquid helium and for allowing us to use the laser Raman microscope, respectively.





\bibliography{hex_nanopit}

\clearpage

\section*{Supporting Information}
\beginsupplement

\subsection*{Raman Spectroscopy of Multilayer Graphene Etched by H-Plasma}

Raman spectroscopy measurements were carried out to cross-check the conclusion drawn from the scanning tunneling spectroscopy (STS) measurements.
We need to use sufficiently thin samples (less than 10 layers) otherwise surface information can easily be masked by bulk information since Raman spectroscopy detects molecular vibrations within a few $\mu$m depth from the surface.
Such multilayer graphene samples were prepared by exfoliating graphite on a silicon wafer which has a 285~nm thick $\mathrm{SiO_2}$ toplayer.

Figure~\ref{fig:multilayer}a is an optical microscope image of such a multilayer graphene flake (Sample A) obtained by H-plasma etching under a low H$_2$ pressure of 13~Pa at $T = 600~^{\circ}\mathrm{C}$ and $W_\mathrm{RF} = 20$~W for 15~min, hereafter the LP etching.
LP etching creates multi-layer terraced nanopits with high densities. 
Figure~\ref{fig:multilayer}b is an atomic force microscope (AFM) image of the region indicated by the black square and arrow in Figure~\ref{fig:multilayer}a.
The maximum thickness of this region is 7 layers or 2.5~nm.
The heavily etched surface is very close to the graphite surface etched under similar conditions (see Figure~1a). 

By applying successive H-plasma etching at a middle pressure of $190~\mathrm{Pa}$ at $T = 600~^{\circ}\mathrm{C}$ and $W_\mathrm{RF} = 20$~W, hereafter the MP etching, following the LP etching, we were able to smoothen the rough surface to a flat one with few large hexagonal nanopits of monolayer depth.
Figure~\ref{fig:multilayer}c is an optical microscope image of a sample obtained by MP etching for 30~min.~following LP one for 15~min.~(Sample B).
The maximum thickness here is 8 layers or 2.7~nm. 
Figure~{\ref{fig:multilayer}d is an AFM image of the region indicated in (c), where large and well shaped hexagonal nanopits are created as expected.
It is interesting to note that the nanopit size is rather uniform here compared to that in graphite (Figure~1f).

Figure~\ref{fig:multilayer}e shows Raman spectra obtained at the same regions where the AFM images were taken for Sample A and B.
Clearly, a significant D band peak appears in Sample A indicating an irregular edge structure (the blue dots and line in the figure).
The D band peak originates from intervalley scattering at irregularly shaped edges, including armchair edges~\cite{You2008,Casiraghi2009}.
On the other hand, in Sample B, it almost disappears indicating high-purity zigzag edges around the hexagonal nanopits recovered by the successive MP etching (the red dots and line), even though there still remain many edges.
This supports the conclusion drawn from the STS measurements.
By taking a more detailed look, it can be seen that the D band peak intensity does not return exactly to the background intensity, presumably due to remnant irregularities of the edges. 

The present Raman results are essentially the same as those obtained by the layer-by-layer thinning technique developed by Yang et al.~\cite{Yang2014} where a mild oxygen plasma etching at room temperature was employed instead of the LP H-plasma etching used here. 

\begin{figure}[H]
\centering
\includegraphics[width=0.8\linewidth]{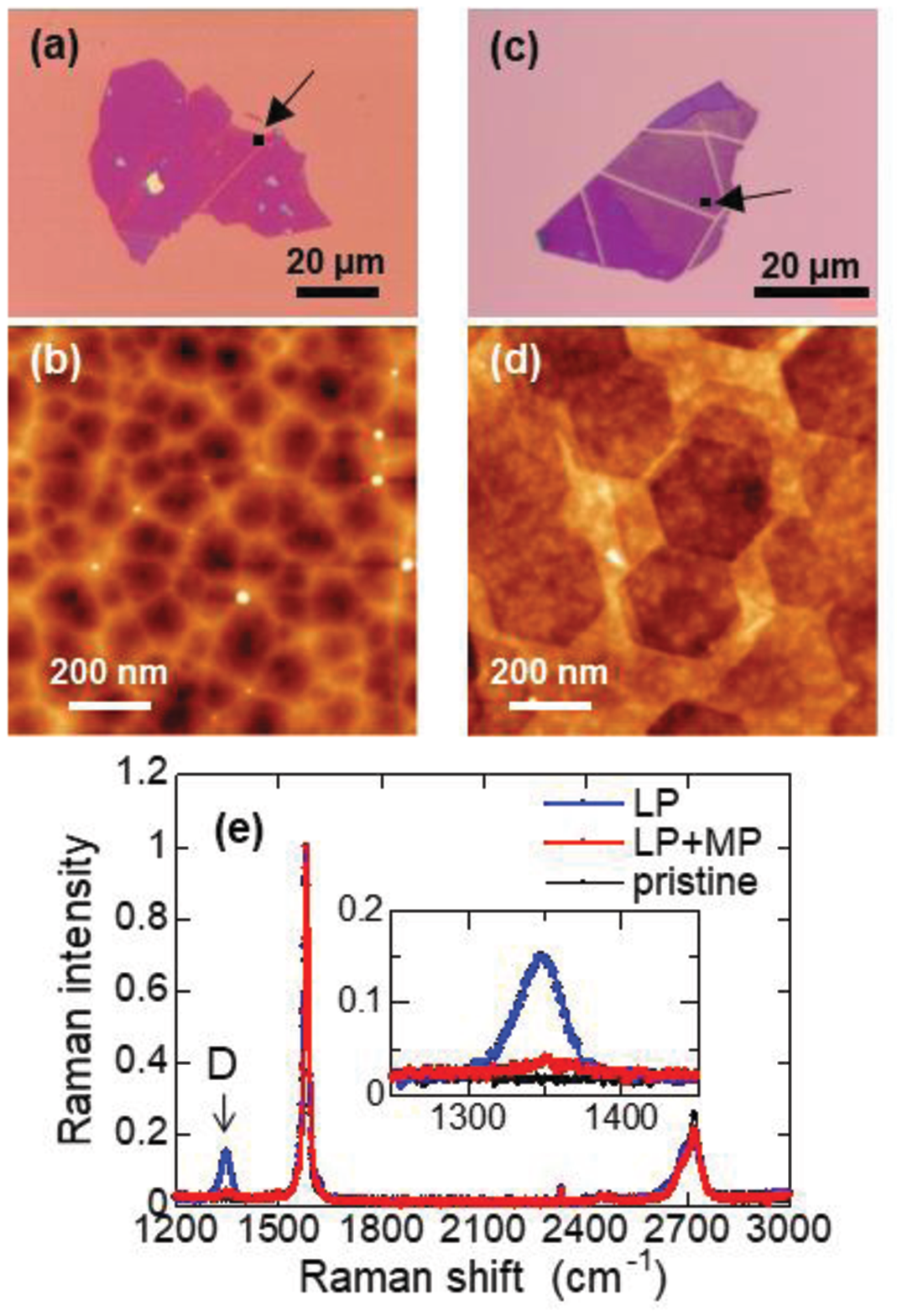}
\caption{
(a) Optical microscope image of a multilayer graphene flake (Sample A) etched by H-plasma at $T = 600~^\circ\mathrm{C}$, $P = 13$~Pa, $t =15$~min, and $W_\mathrm{RF} = 20$~W (LP etching). 
(b) AFM image of the region indicated by the square and arrow in (a). 
(c) Optical microscope image of a multilayer graphene flake (Sample B) etched at $T = 600~^\circ\mathrm{C}$, $P = 190$~Pa, $t =30$~min, and $W_\mathrm{RF} = 20$~W (MP etching) after LP etching. 
(d) AFM image of the region indicated by the square and arrow in (c). 
(e) Raman spectra for Sample A (blue: region (b)), Sample B (red: region (d)), and an as-exfoliated graphite flake (black). 
Inset: Magnification of the spectra around the D band peak. 
All the samples were exfoliated on SiO$_2$/Si substrates.
}
\label{fig:multilayer}
\end{figure}

\end{document}